\documentclass[aps,prl,onecolumn,floatfix,superscriptaddress,reprint,longbibliography]{revtex4-1}
\usepackage[T1]{fontenc}
\usepackage[utf8]{inputenc}
\usepackage{graphicx}
\setcitestyle{super}
\usepackage{color}
\usepackage{txfonts}
\usepackage{times}
\usepackage{hyperref}
\usepackage{enumitem}  
\usepackage{booktabs}
\usepackage{tabularx}

\definecolor{myblue}{RGB}{0, 30, 120}
\hypersetup{
    pdftoolbar=true,        % show Acrobat’s toolbar?
    pdfmenubar=true,        % show Acrobat’s menu?
    pdffitwindow=false,     % window fit to page when opened
    pdfstartview={FitH},    % fits the width of the page to the window
    unicode=true,          % non-Latin characters in Acrobat’s 
    pdftitle={Weakly bound molecules as sensors of new gravitylike forces},    
    pdfauthor={Mateusz Borkowski},     
    colorlinks=true,       % false: boxed links; true: colored links
    linkcolor=myblue,          % color of internal links
    citecolor=myblue,        % color of links to bibliography
    filecolor=myblue,      % color of file links
    urlcolor=myblue,           % color of external links
}

\begin{document}
	\title{Weakly bound molecules as sensors of new gravitylike forces}
	
	\author{Mateusz Borkowski}
	\email{mateusz@fizyka.umk.pl}	
	\affiliation{Institute of Physics, Faculty of Physics, Astronomy and Informatics, Nicolaus Copernicus University, Grudziadzka 5, 87-100 Torun, Poland}
	
	\author{Alexei A. Buchachenko}
	\affiliation{Skolkovo Institute of Science and Technology, 100 Novaya Street, Skolkovo, Moscow Region, 143025, Russia}
	
	\author{Roman Ciuryło}
	\affiliation{Institute of Physics, Faculty of Physics, Astronomy and Informatics, Nicolaus Copernicus University, Grudziadzka 5, 87-100 Torun, Poland}
	
	\author{Paul S. Julienne}
	\affiliation{Joint Quantum Institute, NIST and the University of Maryland, College Park, Maryland 20742, USA}
	
	\author{Hirotaka Yamada}
	\author{Yuu Kikuchi}
	\author{Yosuke Takasu}
	\author{Yoshiro Takahashi}
	\affiliation{Department of Physics, Graduate School of Science, Kyoto University, Kyoto 606-8502, Japan}
	
	\date{\today}
	
\maketitle

\textbf{Several extensions to the Standard Model of particle physics, including light dark matter candidates and unification theories \cite{Adelberger2003, Fayet1996,  Adelberger2009, Knapen2017, ArkaniHamed1998, Antoniadis1998}, predict deviations from Newton's law of gravitation. For macroscopic distances, the inverse-square law of gravitation is well confirmed by astrophysical observations and laboratory experiments \cite{Adelberger2003, Adelberger2009, Su1994, Kapner2007, Geraci2008}. At micrometer and shorter length scales, however, even the state-of-the-art constraints on deviations from gravitational interaction, whether provided by neutron scattering \cite{Pokotilovski2006, Nesvizhevsky2008, Kamiya2015} or precise measurements of forces between macroscopic bodies \cite{Klimchitskaya2013, Chen2016}, are currently many orders of magnitude larger than gravity itself. Here we show that precision spectroscopy of weakly bound molecules~\cite{Jones2006} can be used to constrain non-Newtonian interactions between atoms. A proof-of-principle demonstration using recent data from photoassociation spectroscopy of weakly bound Yb$_2$ molecules \cite{Borkowski2017a} yields constraints on these new interactions that are already close to state-of-the-art neutron scattering experiments~\cite{Kamiya2015}. At the same time, with the development of the recently proposed optical molecular clocks \cite{Borkowski2018}, the neutron scattering constraints could be surpassed by at least two orders of magnitude. }

The experimental search for non-Newtonian gravity has been taking place for years~\cite{Adelberger2003}. Experimental bounds on hypothetical nanometer range forces could help verify several extensions to the Standard Model, including grand unification theories~\cite{Fayet1996, Adelberger2009}, light dark matter models~\cite{Knapen2017} and extradimensional theories aimed at solving the hierarchy problem\cite{ArkaniHamed1998,Antoniadis1998,Adelberger2003}. For two atoms with mass numbers $N_{1,2}$, separated by a distance $R$, the existence of a light boson that couples to nucleons would imply an additional Yukawa-type interaction, or ``fifth force''~\cite{Fayet1996, Adelberger2003, Adelberger2009, Knapen2017, Kamiya2015},
\begin{equation}
	V_5(R) = - N_1 N_2 \frac{g^2}{4\pi}  \hbar c \frac{e^{-R/\lambda}}{R}\,, \label{eq:fifthforce}
\end{equation}
whose range $\lambda = \hbar/Mc$ is determined by the mass $M$ of the new particle, while the dimensionless parameter $g^2$ reflects the coupling strength between nucleons and the new particle field. Experimental methods employed to provide bounds on $g^2$ vary greatly depending on the range $\lambda$ of the hypothetical new forces: from astrophysical observations~\cite{Adelberger2003}, to torsion balance experiments \cite{Su1994, Bordag2001, Masuda2009, Sushkov2011, Adelberger2009}, Casimir-less techniques~\cite{Decca2005, Chen2016}, atomic force microscopy~\cite{Klimchitskaya2013}, and finally neutron scattering on a neutral atom target~\cite{Pokotilovski2006, Nesvizhevsky2008, Kamiya2015}. A promising technique based on direct comparison of spectroscopic measurements of deeply bound hydrogen molecules with precise \emph{ab initio} calculations was also demonstrated\cite{Salumbides2013}. While stringent for macroscopic ranges $\lambda$, the experimental constraints on corrections to gravity for tens of micrometers or less quickly become many orders of magnitude larger than gravity itself due to the presence of much stronger van der Waals or Casimir interactions at these length scales~\cite{Kapner2007,Chen2016}.

\begin{figure}[t]
\centering
	\includegraphics[width=0.46\textwidth]{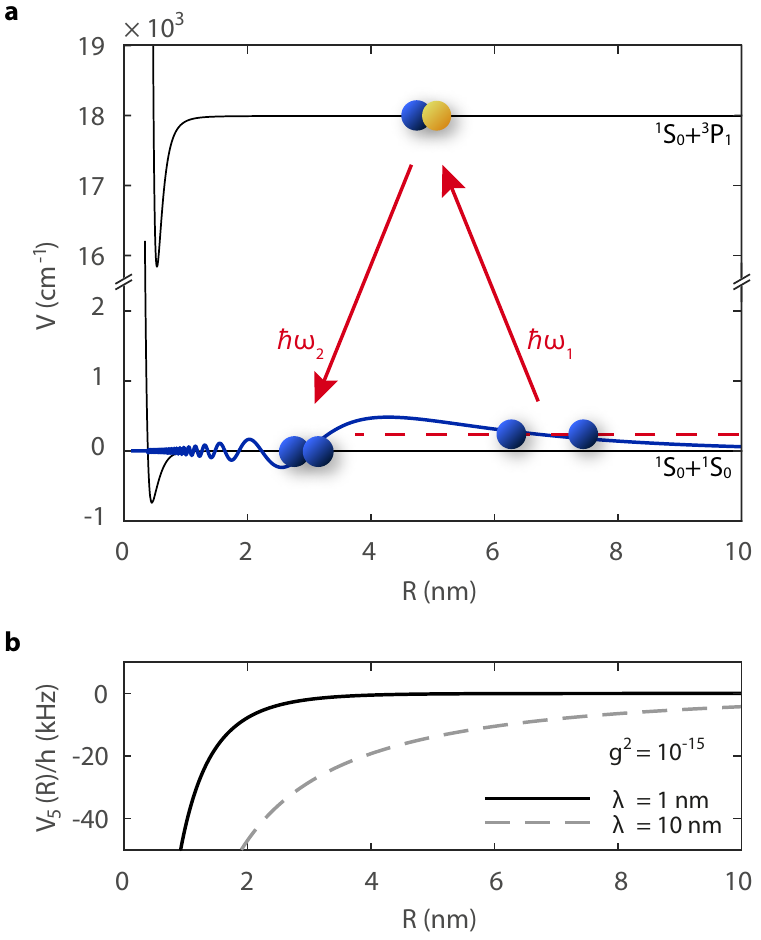}
	%\vspace{11 cm}
	\caption{\textbf{New gravitylike forces and long-range atomic interactions.} \textbf{a}, A schematic depiction of the principle of two color photoassociation spectroscopy. The vibrational wavefunction for a $^{170}$Yb$_2$ bound state with a vibrational quantum number $v'=-2$ (as counted from the dissociation limit) and total angular momentum $J=0$ at a binding energy $E_b/h=-463.72552(80)$~MHz peaks at $R\approx 4.2$~nm (blue). \textbf{b}, Example  Yukawa-type gravitylike potentials $V_5(R)$ for two $^{170}$Yb atoms, as defined by Eq.~(\ref{eq:fifthforce}). \label{fig1}}
\end{figure}

Here we propose to search for new gravitylike forces in long range atomic interactions using high precision spectroscopy of weakly bound ultracold molecules. Unlike deeply bound hydrogen dimers~\cite{Salumbides2013}, where the equilibrium distance lies at $R\approx 0.074$~nm, the vibrational motion in bound states close to a molecule's dissociation limit can extend to several nanometers (Fig.~\ref{fig1}a)\cite{Jones2006}.  The presence of a new Yukawa-type potential could be manifested as a perturbation to near-threshold vibrational series. For $g^2=10^{-15}$, comparable to current limits in the nanometer range~\cite{Kamiya2015}, $V_5(R)$ could contribute an additional several to several tens of kHz to the interaction potential for $R$ comparable to the size of the molecule (Fig.\ref{fig1}b). Weakly bound molecules composed of bosonic two-valence-electron atoms, like Yb, Sr, or Hg, lend themselves to simple theoretical descriptions thanks to their spin-singlet electronic ground state and a lack of hyperfine structure. Near-threshold vibrational splittings depend chiefly on the dominant long range $R^{-6}$ van der Waals interaction and are to a large extent insensitive to the details of the short range potential~\cite{Leroy1970, Jones2006}. The narrow intercombination lines present in divalent species facilitate measurements of the positions of near-threshold bound states of Yb$_2$~\cite{Borkowski2017a} and Sr$_2$~\cite{Stellmer2012, McGuyer2015a} to an already impressive sub-kHz accuracy which in the future could further be improved by several orders of magnitude using lattice clock techniques~\cite{Borkowski2018}. Thus, weakly bound molecules composed of Yb or Sr atoms make excellent testing grounds in the search for new interactions by uniting precision measurements with a relatively simple theoretical description. 

\begin{table}[b]
	\caption{\textbf{Vibrational state positions for ground state Yb$_2$ molecules~\cite{Borkowski2017a}.} All bound state positions are given in MHz with respect to the $^1$S$_0$+$^1$S$_0$ dissociation limit. The quantum numbers $v'$ and $J$ correspond to, respectively, the vibrational quantum number (counted from the dissociation limit), and the total angular momentum. Values in parentheses are standard uncertainties. \label{table1}}
	\centering
	\begin{tabularx}{0.47\textwidth}{r r r r r}
	    \toprule
		$v'$ & $J$ &       $^{168}$Yb &        $^{170}$Yb & $^{174}$Yb \\
		\hline
		$-1$ & $2$ & 			      &    $-3.66831(32)$ &            \\
		$-1$ & $0$ & 		          &   $-27.70024(44)$ &   $-10.62513(53)$ \\
		$-2$ & $2$ & $-145.53196(48)$ &  $-398.05626(46)$ &  $-268.63656(56)$ \\
		$-2$ & $0$ & $-195.18141(46)$ &  $-463.72552(80)$ &  $-325.66378(98)$ \\
		$-3$ & $2$ & 			      & $-1817.14074(80)$ & $-1432.82653(75)$ \\
		$-3$ & $0$ & 			     & $-1922.01467(505)$ & $-1527.88543(34)$ \\
		\toprule
	\end{tabularx}
\end{table}

We demonstrate our proposal by carrying out a proof-of-concept determination of constraints on the new forces using the recent state-of-the-art measurements of near-threshold Yb$_2$ bound state energies~\cite{Borkowski2017a}.  With a total of 13 rovibrational state positions (Table~\ref{table1}) for three Yb$_2$ isotopomers it is the largest of the currently available sub-kHz datasets. The bound state energies were measured using two-color photoassociation spectroscopy~\cite{Jones2006} of Yb Bose-Einstein condensates in optical dipole traps. Here, two lasers were used to induce Raman coupling between colliding atomic pairs and a rovibrational level in the electronic ground state using an intermediate excited state. Once the difference in the laser frequencies $\hbar \omega_1 - \hbar \omega_2$ matched the energy $E_b$ of a vibrational level in the electronic ground state with respect to the dissociation limit, loss of atoms from the trap was observed. Systematic shifts from the trapping and photoassociation lasers, and the mean-field shift of the BECs have been taken into account leading to experimental uncertainties $\approx 500$~Hz for most bound state energies.

\begin{figure}
	\includegraphics[width=0.47\textwidth]{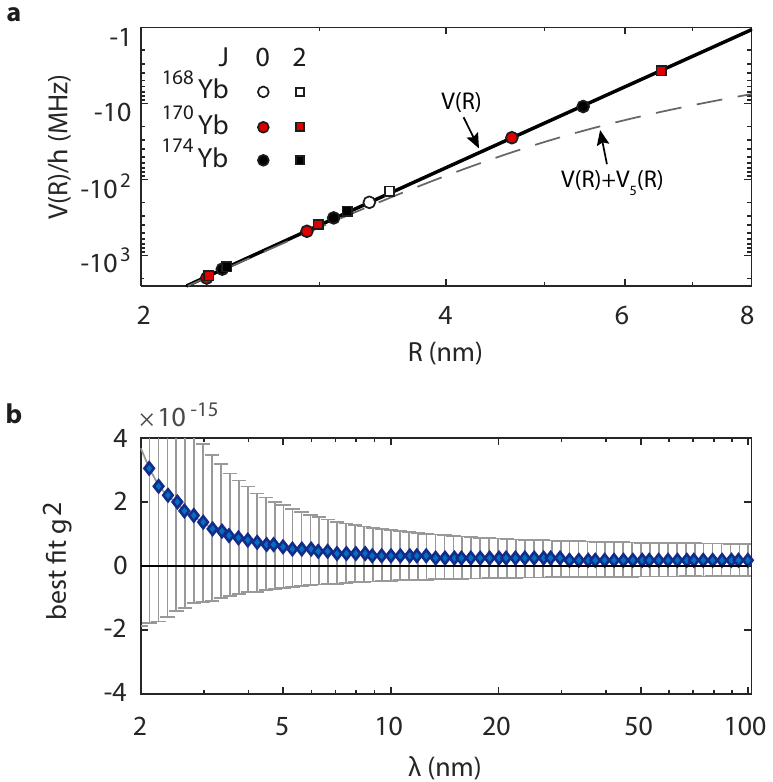}
	\caption{\textbf{Sensitivity of long range atomic interactions to new gravitylike forces.} \textbf{a},~Long range $R^{-6}$ van der Waals interaction between two Yb atoms. Markers indicate positions of bound states measured in photoassociative spectroscopy \cite{Borkowski2017a}. Dashed line shows the same potential modified by an additional Yukawa interaction for $g^2=10^{-13}$ (much larger than current limits at nanometers to make it visible in plot) and $\lambda=1$~nm. \textbf{b},~Best fit coupling parameters $g^2$ as a function of $\lambda$. All of the fitted $g^2$ values are compatible with zero (horizontal line) well within one standard uncertainty (shown as error bars). \label{fig2}}
\end{figure}

The measured binding energy range of $-1922$ to $-3.7$~MHz corresponds to classical outer turning points in between $R=2.3$ and $R=6.5$ nm (Fig.~\ref{fig2}a). At these internuclear distances the atomic potential is dominated by the long range $R^{-6}$ van der Waals interaction.  Adding the Yukawa-type potential $V_5(R)$ imposes a significant change to the long range atomic interaction that can be well distinguished from the expected $R^{-6}$ behavior. We describe the interactions between two Yb atoms using our state-of-the-art mass-scaled interaction model~\cite{Borkowski2017a}. For a total angular momentum $J$ the rovibrational level energies obey the radial Schr\"odinger equation,
\begin{equation}
\left(-\frac{\hbar^2}{2\mu} \frac{d^2}{dR^2} + V(R) + V_{\rm ad}(R)+\frac{\hbar^2 J(J+1)}{2\mu R^2} \right) \Psi(R)  =  E_b  \Psi(R) \,. \label{eq:radial}
\end{equation}  
Since both atoms are in structureless $^1$S$_0$ electronic ground states, there are no permanent multipole moments and at large separations the atoms interact purely due to dispersion. The long range part of the interaction potential $V(R)$ is dominated by the induced dipole-dipole $C_6R^{-6}$ term, with $C_6\approx 1937\,E_h a_0^6$ (the Hartree energy $E_h\approx4.359744650(54)\times 10^{-18}$~J and Bohr radius $a_0\approx5.2917721067(12)\times10^{-11}$~m are the atomic units of energy and distance). The next dispersion term $C_8R^{-8}$, with $C_8\approx2.265\times10^5\, E_h a_0^8$, describes the induced dipole-quadrupole interaction. Although at $R=5$~nm it represents just 1.3\% of the potential energy, it is critical to reach proper quality of the fit. On the other hand, introducing the next dispersion term, $C_{10}R^{-10}$, fails to improve the fit. Similarly, no improvement is seen when introducing the Casimir-Polder effect, whether by directly implementing \emph{ab initio} corrections to the long range potential~\cite{Zhang2008} or adding a fitted $+w_4R^{-4}$ term. The analytic dispersive interaction $V(R) \rightarrow -C_6R^{-6}-C_8R^{-8}$ is smoothly connected to a realistic \emph{ab initio} short range potential. The best fit potential depth $D_e\approx739.7\,\textrm{cm}^{-1}$ is set by scaling the \emph{ab initio} potential by just 2.3\%. Our interaction model also includes two beyond-Born-Oppenheimer effects -- the adiabatic correction $V_{\rm ad}(R)$ as calculated by Lutz and Hutson~\cite{Lutz2016} and an $R$-dependent effective reduced mass $\mu$~\cite{Pachucki2008}. The latter is a nonadiabatic effect and is modeled by having the reduced mass $\mu$ vary smoothly between half the nuclear mass for $R\to0$ and half the atomic mass when the two atoms are well separated ($R\to\infty$)~\cite{Borkowski2017a}. The parameters $C_6$, $C_8$ and $D_e$ are fitted to the experimental data by nonlinear least squares. The van der Waals parameters $C_6$, and to a lesser extent $C_8$, determine the near-threshold vibrational spacings, whereas the depth $D_e$ fixes the phase of the short range radial wavefunction and, by proxy, the position of the entire near-threshold vibrational spectrum \cite{Leroy1970, Jones2006}. Our interaction model reproduces the positions of near-threshold bound states in the Yb$_2$ molecule to $\approx 30$~kHz.

\begin{figure}[t]
    \centering
    \includegraphics[width=0.47\textwidth]{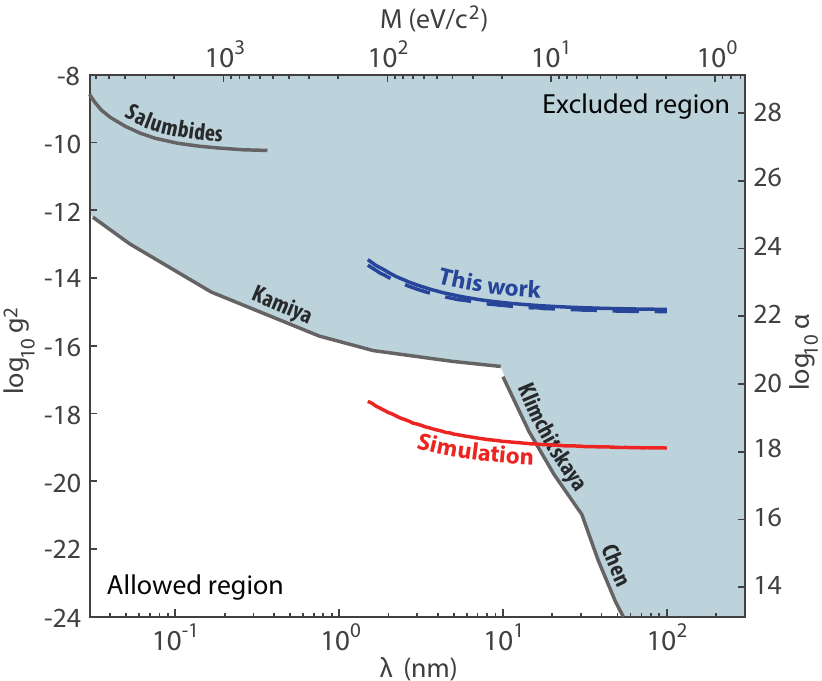}
	\caption{\textbf{Constraints on Yukawa-type gravitylike forces.} Feldman-Cousins limits on $g^2$ as a function of Yukawa range $\lambda=\hbar/Mc$, where $M$ is the mass of the hypothetical new particle, and comparison with other constraints derived from spectroscopy of hydrogen molecules (Salumbides \emph{et al.}~\cite{Salumbides2013}), neutron scattering (Kamiya \emph{et al.}~\cite{Kamiya2015}), atomic force microscopy~(Klimchitskaya \emph{et al.}\cite{Klimchitskaya2013}) and a Casimir-less experiment (Chen \emph{et al.}\cite{Chen2016}). Since the interaction $V_5(R)$ may also be viewed as a correction to the gravitational interaction between two test masses $m_1$ and $m_2$, $V_g(R) = -(Gm_1m_2/R)\left( 1+\alpha\exp(-R/\lambda)\right)$, we also show our constraints in terms of $\alpha \approx (\hbar c)/(4\pi G m_p^2) g^2 \approx 1.347 \times 10^{37} g^2$, where $m_p$ is the proton mass \cite{Fayet1996}. \label{fig3}}
\end{figure}

We first extract our limits on the magnitude of the coupling parameter $g^2$ using the complete dataset. We add the new interaction $V_5(R)$ to the Hamiltonian in Eq.~(\ref{eq:radial}) and run a series of least-squares fits for varying Yukawa ranges $\lambda$. In each fit the $\lambda$ parameter is held fixed, whereas the three adjustable potential parameters, $C_6$, $C_8$, and $D_e$~\cite{Borkowski2017a} and now also the coupling $g^2$, are optimized again using nonlinear least-squares. The uncertainties for the four fitted parameters have been scaled by the factor $\sqrt{\chi^2/\textrm{dof}}$, where $\textrm{dof} = 13-4-1=8$ is the number of degrees of freedom. For our dataset the fits converge reliably for $\lambda = 2\ldots 100$~nm. The resulting $g^2$ values are all compatible with zero well within one standard uncertainty (Fig.~\ref{fig2}b). Following Kamiya~\emph{et al.}\cite{Kamiya2015}, we determine the 95\% confidence limits (Fig.~\ref{fig3}) using the Feldman-Cousins approach\cite{Feldman1998} which takes into account the fact that $g^2$ should have a non-negative value. 
Secondly, we verify that our constraints are due to the impact the Yukawa potential has on long range interactions, rather than its dependence on the number of nucleons. To do so, we have repeated our fitting procedure but with the dataset restricted to $^{170}$Yb$_2$. Only for this isotope a sufficient number of experimental data points is available to allow a convincing fit for four fitted parameters ($\textrm{dof} = 6-4-1=1$). 
Finally, we run a projection for a hypothetical scenario where theory could fit experimental data to within 1~Hz. The state-of-the-art measurements of bound state positions in weakly bound molecules reach an accuracy of hundreds of Hz~\cite{Stellmer2012, McGuyer2015a, Borkowski2017a}, which may in the near future be improved by several orders of magnitude using molecular clock transitions~\cite{Borkowski2018}. Atomic optical clocks currently have short-term relative instabilities of about 10$^{-15}$ ($\sim$1~Hz absolute), and with proper averaging reach a relative accuracy of 10$^{-18}$. Thus, sub-Hz-level measurements of molecular level positions could reasonably be attainable. Conversely, the constraints on new gravitylike forces could improve by several orders of magnitude. To obtain the projected constraint (``Simulation'' in Fig.~\ref{fig3}) we used a simulated dataset, comprised of theoretical bound state positions for the same bound states as listed in Table~\ref{table1}, calculated from Eq.~(\ref{eq:radial}) with an added Gaussian noise with a standard deviation of 1~Hz.

The constraints obtained for the current photoassociative dataset are already close to the current state-of-the-art. For a Yukawa range of $\lambda = 10~\rm{nm}$ the best fit coupling is $g^2 = (3.2 \pm 7.9) \times 10^{-16}$, which corresponds to a Feldman-Cousins 95\% confidence level limit of $g^2 \leq 1.9\times10^{-15}$, just two orders of magnitude above the neutron scattering constraints of Kamiya~\emph{et al.}~\cite{Kamiya2015}. Restricting the dataset to $^{170}$Yb$_2$ results in nearly identical, and even slightly more stringent constraints. This can be explained by the fact that it is easier to accurately reproduce the photoassociation spectra for a single isotope than to construct a fully mass-scaled model~\cite{Borkowski2017a}. When limited to one isotope, the model can fit the photoassociation data to within about 10~kHz, rather than the 30~kHz for a mass-scaled model. This shows that it may be a better strategy to measure many lines for a single isotope (e.g. for many rotational levels) rather than use many isotopes. At a certain level of accuracy it will be necessary to take into account e.g. the slight isotopic dependence of the van der Waals coefficients or the potential depth and may require separate fitting parameters for each isotope. This problem could be mitigated by calculating the small isotopic differences using \emph{ab initio} methods. The few-percent relative accuracy typical for \emph{ab initio} calculations for heavy dimers may suffice for small corrections.

Our projected constraints for a hypothetical scenario, where theory matches experiment to within 1~Hz, show a significant potential for our method. For instance, the current limits for $\lambda = 2$~nm to $\lambda = 10$~nm could be surpassed by about 1.5--2 orders of magnitude. This, however, will also require the inclusion of several subtle QED~\cite{Zhang2008} and relativistic effects~\cite{Balcerzak2017} in the theoretical description of long range atomic interactions. If data for many isotopes are to be used, an \emph{ab initio} calculation of isotope-dependent corrections, like the adiabatic, nonadiabatic or nuclear volume corrections~\cite{Lutz2016} may prove necessary. In the future, the mass-dependency of the Yukawa gravitylike forces could additionally constrain their magnitude, through its impact on the mass-scaling behavior of the near-threshold bound state positions between different isotopes. Even short range forces, where $\lambda$ is much smaller than the ranges investigated here, could impact the phase of the short range wavefunction in a detectable manner~\cite{Lutz2016}. Such attempts, however, will require a full understanding of the mass-dependent Beyond-Born-Oppenheimer corrections~\cite{Pachucki2008, Lutz2016, Borkowski2017a}.

In conclusion, we have proposed and demonstrated a new method for constraining new Yukawa-type gravitylike forces in the nanometer range based on precision spectroscopy of near-threshold molecular states. Ultracold weakly bound molecules composed of ground state spin-singlet atoms, like Yb or Sr, are an excellent testing ground in searching for new interactions thanks to their simple structure and narrow optical transitions that allow for precision measurements. The available photoassociation data~\cite{Borkowski2017a} for the Yb$_2$ molecule already makes it possible to derive constraints on new nanometer range Yukawa-type forces close to current state-of-the-art constraints derived from mature experimental techniques like neutron scattering~\cite{Kamiya2015} or measurements of Casimir-Polder forces~\cite{Klimchitskaya2013}. Our method is complementary to the spectroscopy of deeply bound hydrogen molecules (Salumbides \emph{et al.}~\cite{Salumbides2013}), as it excels for Yukawa ranges of several nanometers, complementing the range of $\sim{}0.1$~nm probed in the latter. In the future, with the development of the recently proposed optical molecular clocks~\cite{Borkowski2018}, our technique could constrain new gravitylike forces at unprecedented levels and provide a valuable means of testing new physics beyond the Standard Model~\cite{Adelberger2003, Adelberger2009, Fayet1996, Knapen2017,Antoniadis1998, ArkaniHamed1998}. 

\bibliography{library}

\begin{thebibliography}{30}%
\makeatletter
\providecommand \@ifxundefined [1]{%
 \@ifx{#1\undefined}
}%
\providecommand \@ifnum [1]{%
 \ifnum #1\expandafter \@firstoftwo
 \else \expandafter \@secondoftwo
 \fi
}%
\providecommand \@ifx [1]{%
 \ifx #1\expandafter \@firstoftwo
 \else \expandafter \@secondoftwo
 \fi
}%
\providecommand \natexlab [1]{#1}%
\providecommand \enquote  [1]{``#1''}%
\providecommand \bibnamefont  [1]{#1}%
\providecommand \bibfnamefont [1]{#1}%
\providecommand \citenamefont [1]{#1}%
\providecommand \href@noop [0]{\@secondoftwo}%
\providecommand \href [0]{\begingroup \@sanitize@url \@href}%
\providecommand \@href[1]{\@@startlink{#1}\@@href}%
\providecommand \@@href[1]{\endgroup#1\@@endlink}%
\providecommand \@sanitize@url [0]{\catcode `\\12\catcode `\$12\catcode
  `\&12\catcode `\#12\catcode `\^12\catcode `\_12\catcode `\%12\relax}%
\providecommand \@@startlink[1]{}%
\providecommand \@@endlink[0]{}%
\providecommand \url  [0]{\begingroup\@sanitize@url \@url }%
\providecommand \@url [1]{\endgroup\@href {#1}{\urlprefix }}%
\providecommand \urlprefix  [0]{URL }%
\providecommand \Eprint [0]{\href }%
\providecommand \doibase [0]{http://dx.doi.org/}%
\providecommand \selectlanguage [0]{\@gobble}%
\providecommand \bibinfo  [0]{\@secondoftwo}%
\providecommand \bibfield  [0]{\@secondoftwo}%
\providecommand \translation [1]{[#1]}%
\providecommand \BibitemOpen [0]{}%
\providecommand \bibitemStop [0]{}%
\providecommand \bibitemNoStop [0]{.\EOS\space}%
\providecommand \EOS [0]{\spacefactor3000\relax}%
\providecommand \BibitemShut  [1]{\csname bibitem#1\endcsname}%
\let\auto@bib@innerbib\@empty
%</preamble>
\bibitem [{\citenamefont {Adelberger}\ \emph {et~al.}(2003)\citenamefont
  {Adelberger}, \citenamefont {Heckel},\ and\ \citenamefont
  {Nelson}}]{Adelberger2003}%
  \BibitemOpen
  \bibfield  {author} {\bibinfo {author} {\bibfnamefont {E.~G.}\ \bibnamefont
  {Adelberger}}, \bibinfo {author} {\bibfnamefont {B.~R.}\ \bibnamefont
  {Heckel}}, \ and\ \bibinfo {author} {\bibfnamefont {A.~E.}\ \bibnamefont
  {Nelson}},\ }\bibfield  {title} {\enquote {\bibinfo {title} {{Tests of the
  Gravitational Inverse-Square Law}},}\ }\href {\doibase
  10.1146/annurev.nucl.53.041002.110503} {\bibfield  {journal} {\bibinfo
  {journal} {Annual Review of Nuclear and Particle Science}\ }\textbf {\bibinfo
  {volume} {53}},\ \bibinfo {pages} {77--121} (\bibinfo {year}
  {2003})}\BibitemShut {NoStop}%
\bibitem [{\citenamefont {Fayet}(1996)}]{Fayet1996}%
  \BibitemOpen
  \bibfield  {author} {\bibinfo {author} {\bibfnamefont {P.}~\bibnamefont
  {Fayet}},\ }\bibfield  {title} {\enquote {\bibinfo {title} {{New interactions
  and the standard models}},}\ }\href {\doibase 10.1088/0264-9381/13/11A/004}
  {\bibfield  {journal} {\bibinfo  {journal} {Classical and Quantum Gravity}\
  }\textbf {\bibinfo {volume} {13}},\ \bibinfo {pages} {A19--A31} (\bibinfo
  {year} {1996})}\BibitemShut {NoStop}%
\bibitem [{\citenamefont {Adelberger}\ \emph {et~al.}(2009)\citenamefont
  {Adelberger}, \citenamefont {Gundlach}, \citenamefont {Heckel}, \citenamefont
  {Hoedl},\ and\ \citenamefont {Schlamminger}}]{Adelberger2009}%
  \BibitemOpen
  \bibfield  {author} {\bibinfo {author} {\bibfnamefont {E.~G.}\ \bibnamefont
  {Adelberger}}, \bibinfo {author} {\bibfnamefont {J.~H.}\ \bibnamefont
  {Gundlach}}, \bibinfo {author} {\bibfnamefont {B.~R.}\ \bibnamefont
  {Heckel}}, \bibinfo {author} {\bibfnamefont {S.}~\bibnamefont {Hoedl}}, \
  and\ \bibinfo {author} {\bibfnamefont {S.}~\bibnamefont {Schlamminger}},\
  }\bibfield  {title} {\enquote {\bibinfo {title} {{Torsion balance
  experiments: A low-energy frontier of particle physics}},}\ }\href {\doibase
  10.1016/j.ppnp.2008.08.002} {\bibfield  {journal} {\bibinfo  {journal}
  {Progress in Particle and Nuclear Physics}\ }\textbf {\bibinfo {volume}
  {62}},\ \bibinfo {pages} {102--134} (\bibinfo {year} {2009})}\BibitemShut
  {NoStop}%
\bibitem [{\citenamefont {Knapen}\ \emph {et~al.}(2017)\citenamefont {Knapen},
  \citenamefont {Lin},\ and\ \citenamefont {Zurek}}]{Knapen2017}%
  \BibitemOpen
  \bibfield  {author} {\bibinfo {author} {\bibfnamefont {S.}~\bibnamefont
  {Knapen}}, \bibinfo {author} {\bibfnamefont {T.}~\bibnamefont {Lin}}, \ and\
  \bibinfo {author} {\bibfnamefont {K.~M.}\ \bibnamefont {Zurek}},\ }\bibfield
  {title} {\enquote {\bibinfo {title} {{Light dark matter: Models and
  constraints}},}\ }\href {\doibase 10.1103/PhysRevD.96.115021} {\bibfield
  {journal} {\bibinfo  {journal} {Physical Review D}\ }\textbf {\bibinfo
  {volume} {96}},\ \bibinfo {pages} {115021} (\bibinfo {year}
  {2017})}\BibitemShut {NoStop}%
\bibitem [{\citenamefont {Arkani-Hamed}\ \emph {et~al.}(1998)\citenamefont
  {Arkani-Hamed}, \citenamefont {Dimopoulos},\ and\ \citenamefont
  {Dvali}}]{ArkaniHamed1998}%
  \BibitemOpen
  \bibfield  {author} {\bibinfo {author} {\bibfnamefont {N.}~\bibnamefont
  {Arkani-Hamed}}, \bibinfo {author} {\bibfnamefont {S.}~\bibnamefont
  {Dimopoulos}}, \ and\ \bibinfo {author} {\bibfnamefont {G.}~\bibnamefont
  {Dvali}},\ }\bibfield  {title} {\enquote {\bibinfo {title} {{The hierarchy
  problem and new dimensions at a millimeter}},}\ }\href {\doibase
  10.1016/S0370-2693(98)00466-3} {\bibfield  {journal} {\bibinfo  {journal}
  {Physics Letters B}\ }\textbf {\bibinfo {volume} {429}},\ \bibinfo {pages}
  {263--272} (\bibinfo {year} {1998})}\BibitemShut {NoStop}%
\bibitem [{\citenamefont {Antoniadis}\ \emph {et~al.}(1998)\citenamefont
  {Antoniadis}, \citenamefont {Arkani-Hamed}, \citenamefont {Dimopoulos},\ and\
  \citenamefont {Dvali}}]{Antoniadis1998}%
  \BibitemOpen
  \bibfield  {author} {\bibinfo {author} {\bibfnamefont {I.}~\bibnamefont
  {Antoniadis}}, \bibinfo {author} {\bibfnamefont {N.}~\bibnamefont
  {Arkani-Hamed}}, \bibinfo {author} {\bibfnamefont {S.}~\bibnamefont
  {Dimopoulos}}, \ and\ \bibinfo {author} {\bibfnamefont {G.}~\bibnamefont
  {Dvali}},\ }\bibfield  {title} {\enquote {\bibinfo {title} {{New dimensions
  at a millimeter to a fermi and superstrings at a TeV}},}\ }\href {\doibase
  10.1016/S0370-2693(98)00860-0} {\bibfield  {journal} {\bibinfo  {journal}
  {Physics Letters B}\ }\textbf {\bibinfo {volume} {436}},\ \bibinfo {pages}
  {257--263} (\bibinfo {year} {1998})}\BibitemShut {NoStop}%
\bibitem [{\citenamefont {Su}\ \emph {et~al.}(1994)\citenamefont {Su},
  \citenamefont {Heckel}, \citenamefont {Adelberger}, \citenamefont {Gundlach},
  \citenamefont {Harris}, \citenamefont {Smith},\ and\ \citenamefont
  {Swanson}}]{Su1994}%
  \BibitemOpen
  \bibfield  {author} {\bibinfo {author} {\bibfnamefont {Y.}~\bibnamefont
  {Su}}, \bibinfo {author} {\bibfnamefont {B.~R.}\ \bibnamefont {Heckel}},
  \bibinfo {author} {\bibfnamefont {E.~G.}\ \bibnamefont {Adelberger}},
  \bibinfo {author} {\bibfnamefont {J.~H.}\ \bibnamefont {Gundlach}}, \bibinfo
  {author} {\bibfnamefont {M.}~\bibnamefont {Harris}}, \bibinfo {author}
  {\bibfnamefont {G.~L.}\ \bibnamefont {Smith}}, \ and\ \bibinfo {author}
  {\bibfnamefont {H.~E.}\ \bibnamefont {Swanson}},\ }\bibfield  {title}
  {\enquote {\bibinfo {title} {{New tests of the universality of free fall}},}\
  }\href {https://journals.aps.org/prd/pdf/10.1103/PhysRevD.50.3614} {\bibfield
   {journal} {\bibinfo  {journal} {Phys. Rev. D}\ }\textbf {\bibinfo {volume}
  {50}},\ \bibinfo {pages} {3614} (\bibinfo {year} {1994})}\BibitemShut
  {NoStop}%
\bibitem [{\citenamefont {Kapner}\ \emph {et~al.}(2007)\citenamefont {Kapner},
  \citenamefont {Cook}, \citenamefont {Adelberger}, \citenamefont {Gundlach},
  \citenamefont {Heckel}, \citenamefont {Hoyle},\ and\ \citenamefont
  {Swanson}}]{Kapner2007}%
  \BibitemOpen
  \bibfield  {author} {\bibinfo {author} {\bibfnamefont {D.~J.}\ \bibnamefont
  {Kapner}}, \bibinfo {author} {\bibfnamefont {T.~S.}\ \bibnamefont {Cook}},
  \bibinfo {author} {\bibfnamefont {E.~G.}\ \bibnamefont {Adelberger}},
  \bibinfo {author} {\bibfnamefont {J.~H.}\ \bibnamefont {Gundlach}}, \bibinfo
  {author} {\bibfnamefont {B.~R.}\ \bibnamefont {Heckel}}, \bibinfo {author}
  {\bibfnamefont {C.~D.}\ \bibnamefont {Hoyle}}, \ and\ \bibinfo {author}
  {\bibfnamefont {H.~E.}\ \bibnamefont {Swanson}},\ }\bibfield  {title}
  {\enquote {\bibinfo {title} {{Tests of the Gravitational Inverse-Square Law
  below the Dark-Energy Length Scale}},}\ }\href {\doibase
  10.1103/PhysRevLett.98.021101} {\bibfield  {journal} {\bibinfo  {journal}
  {Physical Review Letters}\ }\textbf {\bibinfo {volume} {98}},\ \bibinfo
  {pages} {021101} (\bibinfo {year} {2007})}\BibitemShut {NoStop}%
\bibitem [{\citenamefont {Geraci}\ \emph {et~al.}(2008)\citenamefont {Geraci},
  \citenamefont {Smullin}, \citenamefont {Weld}, \citenamefont {Chiaverini},\
  and\ \citenamefont {Kapitulnik}}]{Geraci2008}%
  \BibitemOpen
  \bibfield  {author} {\bibinfo {author} {\bibfnamefont {A.~A.}\ \bibnamefont
  {Geraci}}, \bibinfo {author} {\bibfnamefont {S.~J.}\ \bibnamefont {Smullin}},
  \bibinfo {author} {\bibfnamefont {D.~M.}\ \bibnamefont {Weld}}, \bibinfo
  {author} {\bibfnamefont {J.}~\bibnamefont {Chiaverini}}, \ and\ \bibinfo
  {author} {\bibfnamefont {A.}~\bibnamefont {Kapitulnik}},\ }\bibfield  {title}
  {\enquote {\bibinfo {title} {{Improved constraints on non-Newtonian forces at
  10 microns}},}\ }\href {\doibase 10.1103/PhysRevD.78.022002} {\bibfield
  {journal} {\bibinfo  {journal} {Physical Review D}\ }\textbf {\bibinfo
  {volume} {78}},\ \bibinfo {pages} {022002} (\bibinfo {year}
  {2008})}\BibitemShut {NoStop}%
\bibitem [{\citenamefont {Pokotilovski}(2006)}]{Pokotilovski2006}%
  \BibitemOpen
  \bibfield  {author} {\bibinfo {author} {\bibfnamefont {Yu.~N.}\ \bibnamefont
  {Pokotilovski}},\ }\bibfield  {title} {\enquote {\bibinfo {title}
  {{Constraints on new interactions from neutron scattering experiments}},}\
  }\href {\doibase 10.1134/S1063778806060020} {\bibfield  {journal} {\bibinfo
  {journal} {Physics of Atomic Nuclei}\ }\textbf {\bibinfo {volume} {69}},\
  \bibinfo {pages} {924--931} (\bibinfo {year} {2006})}\BibitemShut {NoStop}%
\bibitem [{\citenamefont {Nesvizhevsky}\ \emph {et~al.}(2008)\citenamefont
  {Nesvizhevsky}, \citenamefont {Pignol},\ and\ \citenamefont
  {Protasov}}]{Nesvizhevsky2008}%
  \BibitemOpen
  \bibfield  {author} {\bibinfo {author} {\bibfnamefont {V.~V.}\ \bibnamefont
  {Nesvizhevsky}}, \bibinfo {author} {\bibfnamefont {G.}~\bibnamefont
  {Pignol}}, \ and\ \bibinfo {author} {\bibfnamefont {K.~V.}\ \bibnamefont
  {Protasov}},\ }\bibfield  {title} {\enquote {\bibinfo {title} {{Neutron
  scattering and extra-short-range interactions}},}\ }\href {\doibase
  10.1103/PhysRevD.77.034020} {\bibfield  {journal} {\bibinfo  {journal}
  {Physical Review D}\ }\textbf {\bibinfo {volume} {77}},\ \bibinfo {pages}
  {034020} (\bibinfo {year} {2008})}\BibitemShut {NoStop}%
\bibitem [{\citenamefont {Kamiya}\ \emph {et~al.}(2015)\citenamefont {Kamiya},
  \citenamefont {Itagaki}, \citenamefont {Tani}, \citenamefont {Kim},\ and\
  \citenamefont {Komamiya}}]{Kamiya2015}%
  \BibitemOpen
  \bibfield  {author} {\bibinfo {author} {\bibfnamefont {Y.}~\bibnamefont
  {Kamiya}}, \bibinfo {author} {\bibfnamefont {K.}~\bibnamefont {Itagaki}},
  \bibinfo {author} {\bibfnamefont {M.}~\bibnamefont {Tani}}, \bibinfo {author}
  {\bibfnamefont {G.~N.}\ \bibnamefont {Kim}}, \ and\ \bibinfo {author}
  {\bibfnamefont {S.}~\bibnamefont {Komamiya}},\ }\bibfield  {title} {\enquote
  {\bibinfo {title} {{Constraints on new gravitylike forces in the nanometer
  range}},}\ }\href {\doibase 10.1103/PhysRevLett.114.161101} {\bibfield
  {journal} {\bibinfo  {journal} {Physical Review Letters}\ }\textbf {\bibinfo
  {volume} {114}},\ \bibinfo {pages} {161101} (\bibinfo {year}
  {2015})}\BibitemShut {NoStop}%
\bibitem [{\citenamefont {Klimchitskaya}\ \emph {et~al.}(2013)\citenamefont
  {Klimchitskaya}, \citenamefont {Mohideen},\ and\ \citenamefont
  {Mostepanenko}}]{Klimchitskaya2013}%
  \BibitemOpen
  \bibfield  {author} {\bibinfo {author} {\bibfnamefont {G.~L.}\ \bibnamefont
  {Klimchitskaya}}, \bibinfo {author} {\bibfnamefont {U.}~\bibnamefont
  {Mohideen}}, \ and\ \bibinfo {author} {\bibfnamefont {V.~M.}\ \bibnamefont
  {Mostepanenko}},\ }\bibfield  {title} {\enquote {\bibinfo {title}
  {{Constraints on corrections to Newtonian gravity from two recent
  measurements of the Casimir interaction between metallic surfaces}},}\ }\href
  {\doibase 10.1103/PhysRevD.87.125031} {\bibfield  {journal} {\bibinfo
  {journal} {Physical Review D}\ }\textbf {\bibinfo {volume} {87}},\ \bibinfo
  {pages} {125031} (\bibinfo {year} {2013})}\BibitemShut {NoStop}%
\bibitem [{\citenamefont {Chen}\ \emph {et~al.}(2016)\citenamefont {Chen},
  \citenamefont {Tham}, \citenamefont {Krause}, \citenamefont {L{\'{o}}pez},
  \citenamefont {Fischbach},\ and\ \citenamefont {Decca}}]{Chen2016}%
  \BibitemOpen
  \bibfield  {author} {\bibinfo {author} {\bibfnamefont {Y.-J.}\ \bibnamefont
  {Chen}}, \bibinfo {author} {\bibfnamefont {W.~K.}\ \bibnamefont {Tham}},
  \bibinfo {author} {\bibfnamefont {D.~E.}\ \bibnamefont {Krause}}, \bibinfo
  {author} {\bibfnamefont {D.}~\bibnamefont {L{\'{o}}pez}}, \bibinfo {author}
  {\bibfnamefont {E.}~\bibnamefont {Fischbach}}, \ and\ \bibinfo {author}
  {\bibfnamefont {R.~S.}\ \bibnamefont {Decca}},\ }\bibfield  {title} {\enquote
  {\bibinfo {title} {{Stronger Limits on Hypothetical Yukawa Interactions in
  the 30–8000 nm Range}},}\ }\href {\doibase 10.1103/PhysRevLett.116.221102}
  {\bibfield  {journal} {\bibinfo  {journal} {Physical Review Letters}\
  }\textbf {\bibinfo {volume} {116}},\ \bibinfo {pages} {221102} (\bibinfo
  {year} {2016})}\BibitemShut {NoStop}%
\bibitem [{\citenamefont {Jones}\ \emph {et~al.}(2006)\citenamefont {Jones},
  \citenamefont {Tiesinga}, \citenamefont {Lett},\ and\ \citenamefont
  {Julienne}}]{Jones2006}%
  \BibitemOpen
  \bibfield  {author} {\bibinfo {author} {\bibfnamefont {K.~M.}\ \bibnamefont
  {Jones}}, \bibinfo {author} {\bibfnamefont {E.}~\bibnamefont {Tiesinga}},
  \bibinfo {author} {\bibfnamefont {P.~D.}\ \bibnamefont {Lett}}, \ and\
  \bibinfo {author} {\bibfnamefont {P.~S.}\ \bibnamefont {Julienne}},\
  }\bibfield  {title} {\enquote {\bibinfo {title} {{Ultracold photoassociation
  spectroscopy: Long-range molecules and atomic scattering}},}\ }\href
  {\doibase 10.1103/RevModPhys.78.483} {\bibfield  {journal} {\bibinfo
  {journal} {Reviews of Modern Physics}\ }\textbf {\bibinfo {volume} {78}},\
  \bibinfo {pages} {483--535} (\bibinfo {year} {2006})}\BibitemShut {NoStop}%
\bibitem [{\citenamefont {Borkowski}\ \emph {et~al.}(2017)\citenamefont
  {Borkowski}, \citenamefont {Buchachenko}, \citenamefont {Ciury{\l}o},
  \citenamefont {Julienne}, \citenamefont {Yamada}, \citenamefont {Kikuchi},
  \citenamefont {Takahashi}, \citenamefont {Takasu},\ and\ \citenamefont
  {Takahashi}}]{Borkowski2017a}%
  \BibitemOpen
  \bibfield  {author} {\bibinfo {author} {\bibfnamefont {M.}~\bibnamefont
  {Borkowski}}, \bibinfo {author} {\bibfnamefont {A.~A.}\ \bibnamefont
  {Buchachenko}}, \bibinfo {author} {\bibfnamefont {R.}~\bibnamefont
  {Ciury{\l}o}}, \bibinfo {author} {\bibfnamefont {P.~S.}\ \bibnamefont
  {Julienne}}, \bibinfo {author} {\bibfnamefont {H.}~\bibnamefont {Yamada}},
  \bibinfo {author} {\bibfnamefont {Y.}~\bibnamefont {Kikuchi}}, \bibinfo
  {author} {\bibfnamefont {K.}~\bibnamefont {Takahashi}}, \bibinfo {author}
  {\bibfnamefont {Y.}~\bibnamefont {Takasu}}, \ and\ \bibinfo {author}
  {\bibfnamefont {Y.}~\bibnamefont {Takahashi}},\ }\bibfield  {title} {\enquote
  {\bibinfo {title} {{Beyond-Born-Oppenheimer effects in sub-kHz-precision
  photoassociation spectroscopy of ytterbium atoms}},}\ }\href {\doibase
  10.1103/PhysRevA.96.063405} {\bibfield  {journal} {\bibinfo  {journal} {Phys.
  Rev. A}\ }\textbf {\bibinfo {volume} {96}},\ \bibinfo {pages} {063405}
  (\bibinfo {year} {2017})}\BibitemShut {NoStop}%
\bibitem [{\citenamefont {Borkowski}(2018)}]{Borkowski2018}%
  \BibitemOpen
  \bibfield  {author} {\bibinfo {author} {\bibfnamefont {M.}~\bibnamefont
  {Borkowski}},\ }\bibfield  {title} {\enquote {\bibinfo {title} {{Optical
  Lattice Clocks with Weakly Bound Molecules}},}\ }\href {\doibase
  10.1103/PhysRevLett.120.083202} {\bibfield  {journal} {\bibinfo  {journal}
  {Physical Review Letters}\ }\textbf {\bibinfo {volume} {120}},\ \bibinfo
  {pages} {083202} (\bibinfo {year} {2018})}\BibitemShut {NoStop}%
\bibitem [{\citenamefont {Bordag}\ \emph {et~al.}(2001)\citenamefont {Bordag},
  \citenamefont {Mohideen},\ and\ \citenamefont {Mostepanenko}}]{Bordag2001}%
  \BibitemOpen
  \bibfield  {author} {\bibinfo {author} {\bibfnamefont {M.}~\bibnamefont
  {Bordag}}, \bibinfo {author} {\bibfnamefont {U.}~\bibnamefont {Mohideen}}, \
  and\ \bibinfo {author} {\bibfnamefont {V.~M.}\ \bibnamefont {Mostepanenko}},\
  }\bibfield  {title} {\enquote {\bibinfo {title} {{New developments in the
  Casimir effect}},}\ }\href {\doibase 10.1016/S0370-1573(01)00015-1}
  {\bibfield  {journal} {\bibinfo  {journal} {Physics Reports}\ }\textbf
  {\bibinfo {volume} {353}},\ \bibinfo {pages} {1--205} (\bibinfo {year}
  {2001})}\BibitemShut {NoStop}%
\bibitem [{\citenamefont {Masuda}\ and\ \citenamefont
  {Sasaki}(2009)}]{Masuda2009}%
  \BibitemOpen
  \bibfield  {author} {\bibinfo {author} {\bibfnamefont {M.}~\bibnamefont
  {Masuda}}\ and\ \bibinfo {author} {\bibfnamefont {M.}~\bibnamefont
  {Sasaki}},\ }\bibfield  {title} {\enquote {\bibinfo {title} {{Limits on
  Nonstandard Forces in the Submicrometer Range}},}\ }\href {\doibase
  10.1103/PhysRevLett.102.171101} {\bibfield  {journal} {\bibinfo  {journal}
  {Physical Review Letters}\ }\textbf {\bibinfo {volume} {102}},\ \bibinfo
  {pages} {171101} (\bibinfo {year} {2009})}\BibitemShut {NoStop}%
\bibitem [{\citenamefont {Sushkov}\ \emph {et~al.}(2011)\citenamefont
  {Sushkov}, \citenamefont {Kim}, \citenamefont {Dalvit},\ and\ \citenamefont
  {Lamoreaux}}]{Sushkov2011}%
  \BibitemOpen
  \bibfield  {author} {\bibinfo {author} {\bibfnamefont {A.~O.}\ \bibnamefont
  {Sushkov}}, \bibinfo {author} {\bibfnamefont {W.~J.}\ \bibnamefont {Kim}},
  \bibinfo {author} {\bibfnamefont {D.~A.~R.}\ \bibnamefont {Dalvit}}, \ and\
  \bibinfo {author} {\bibfnamefont {S.~K.}\ \bibnamefont {Lamoreaux}},\
  }\bibfield  {title} {\enquote {\bibinfo {title} {{New Experimental Limits on
  Non-Newtonian Forces in the Micrometer Range}},}\ }\href {\doibase
  10.1103/PhysRevLett.107.171101} {\bibfield  {journal} {\bibinfo  {journal}
  {Physical Review Letters}\ }\textbf {\bibinfo {volume} {107}},\ \bibinfo
  {pages} {171101} (\bibinfo {year} {2011})}\BibitemShut {NoStop}%
\bibitem [{\citenamefont {Decca}\ \emph {et~al.}(2005)\citenamefont {Decca},
  \citenamefont {L{\'{o}}pez}, \citenamefont {Chan}, \citenamefont {Fischbach},
  \citenamefont {Krause},\ and\ \citenamefont {Jamell}}]{Decca2005}%
  \BibitemOpen
  \bibfield  {author} {\bibinfo {author} {\bibfnamefont {R.~S.}\ \bibnamefont
  {Decca}}, \bibinfo {author} {\bibfnamefont {D.}~\bibnamefont {L{\'{o}}pez}},
  \bibinfo {author} {\bibfnamefont {H.~B.}\ \bibnamefont {Chan}}, \bibinfo
  {author} {\bibfnamefont {E.}~\bibnamefont {Fischbach}}, \bibinfo {author}
  {\bibfnamefont {D.~E.}\ \bibnamefont {Krause}}, \ and\ \bibinfo {author}
  {\bibfnamefont {C.~R.}\ \bibnamefont {Jamell}},\ }\bibfield  {title}
  {\enquote {\bibinfo {title} {{Constraining New Forces in the Casimir Regime
  Using the Isoelectronic Technique}},}\ }\href {\doibase
  10.1103/PhysRevLett.94.240401} {\bibfield  {journal} {\bibinfo  {journal}
  {Physical Review Letters}\ }\textbf {\bibinfo {volume} {94}},\ \bibinfo
  {pages} {240401} (\bibinfo {year} {2005})}\BibitemShut {NoStop}%
\bibitem [{\citenamefont {Salumbides}\ \emph {et~al.}(2013)\citenamefont
  {Salumbides}, \citenamefont {Koelemeij}, \citenamefont {Komasa},
  \citenamefont {Pachucki}, \citenamefont {Eikema},\ and\ \citenamefont
  {Ubachs}}]{Salumbides2013}%
  \BibitemOpen
  \bibfield  {author} {\bibinfo {author} {\bibfnamefont {E.~J.}\ \bibnamefont
  {Salumbides}}, \bibinfo {author} {\bibfnamefont {J.~C.~J.}\ \bibnamefont
  {Koelemeij}}, \bibinfo {author} {\bibfnamefont {J.}~\bibnamefont {Komasa}},
  \bibinfo {author} {\bibfnamefont {K.}~\bibnamefont {Pachucki}}, \bibinfo
  {author} {\bibfnamefont {K.~S.~E.}\ \bibnamefont {Eikema}}, \ and\ \bibinfo
  {author} {\bibfnamefont {W.}~\bibnamefont {Ubachs}},\ }\bibfield  {title}
  {\enquote {\bibinfo {title} {{Bounds on fifth forces from precision
  measurements on molecules}},}\ }\href {\doibase 10.1103/PhysRevD.87.112008}
  {\bibfield  {journal} {\bibinfo  {journal} {Physical Review D}\ }\textbf
  {\bibinfo {volume} {87}},\ \bibinfo {pages} {112008} (\bibinfo {year}
  {2013})}\BibitemShut {NoStop}%
\bibitem [{\citenamefont {{Le Roy}}\ and\ \citenamefont
  {Bernstein}(1970)}]{Leroy1970}%
  \BibitemOpen
  \bibfield  {author} {\bibinfo {author} {\bibfnamefont {R.~J.}\ \bibnamefont
  {{Le Roy}}}\ and\ \bibinfo {author} {\bibfnamefont {R.~B.}\ \bibnamefont
  {Bernstein}},\ }\bibfield  {title} {\enquote {\bibinfo {title} {{Dissociation
  Energy and Long-Range Potential of Diatomic Molecules from Vibrational
  Spacings of Higher Levels}},}\ }\href {\doibase 10.1063/1.1697142} {\bibfield
   {journal} {\bibinfo  {journal} {The Journal of Chemical Physics}\ }\textbf
  {\bibinfo {volume} {52}},\ \bibinfo {pages} {3869} (\bibinfo {year}
  {1970})}\BibitemShut {NoStop}%
\bibitem [{\citenamefont {Stellmer}\ \emph {et~al.}(2012)\citenamefont
  {Stellmer}, \citenamefont {Pasquiou}, \citenamefont {Grimm},\ and\
  \citenamefont {Schreck}}]{Stellmer2012}%
  \BibitemOpen
  \bibfield  {author} {\bibinfo {author} {\bibfnamefont {S.}~\bibnamefont
  {Stellmer}}, \bibinfo {author} {\bibfnamefont {B.}~\bibnamefont {Pasquiou}},
  \bibinfo {author} {\bibfnamefont {R.}~\bibnamefont {Grimm}}, \ and\ \bibinfo
  {author} {\bibfnamefont {F.}~\bibnamefont {Schreck}},\ }\bibfield  {title}
  {\enquote {\bibinfo {title} {{Creation of Ultracold Sr$_2$ Molecules in the
  Electronic Ground State}},}\ }\href {\doibase 10.1103/PhysRevLett.109.115302}
  {\bibfield  {journal} {\bibinfo  {journal} {Physical Review Letters}\
  }\textbf {\bibinfo {volume} {109}},\ \bibinfo {pages} {115302} (\bibinfo
  {year} {2012})}\BibitemShut {NoStop}%
\bibitem [{\citenamefont {McGuyer}\ \emph {et~al.}(2015)\citenamefont
  {McGuyer}, \citenamefont {McDonald}, \citenamefont {Iwata}, \citenamefont
  {Tarallo}, \citenamefont {Grier}, \citenamefont {Apfelbeck},\ and\
  \citenamefont {Zelevinsky}}]{McGuyer2015a}%
  \BibitemOpen
  \bibfield  {author} {\bibinfo {author} {\bibfnamefont {B.~H.}\ \bibnamefont
  {McGuyer}}, \bibinfo {author} {\bibfnamefont {M.}~\bibnamefont {McDonald}},
  \bibinfo {author} {\bibfnamefont {G.~Z.}\ \bibnamefont {Iwata}}, \bibinfo
  {author} {\bibfnamefont {M.~G.}\ \bibnamefont {Tarallo}}, \bibinfo {author}
  {\bibfnamefont {A.~T.}\ \bibnamefont {Grier}}, \bibinfo {author}
  {\bibfnamefont {F.}~\bibnamefont {Apfelbeck}}, \ and\ \bibinfo {author}
  {\bibfnamefont {T.}~\bibnamefont {Zelevinsky}},\ }\bibfield  {title}
  {\enquote {\bibinfo {title} {{High-precision spectroscopy of ultracold
  molecules in an optical lattice}},}\ }\href {\doibase
  10.1088/1367-2630/17/5/055004} {\bibfield  {journal} {\bibinfo  {journal}
  {New Journal of Physics}\ }\textbf {\bibinfo {volume} {17}},\ \bibinfo
  {pages} {055004} (\bibinfo {year} {2015})}\BibitemShut {NoStop}%
\bibitem [{\citenamefont {Zhang}\ and\ \citenamefont
  {Dalgarno}(2008)}]{Zhang2008}%
  \BibitemOpen
  \bibfield  {author} {\bibinfo {author} {\bibfnamefont {P.}~\bibnamefont
  {Zhang}}\ and\ \bibinfo {author} {\bibfnamefont {A.}~\bibnamefont
  {Dalgarno}},\ }\bibfield  {title} {\enquote {\bibinfo {title} {{Long-range
  interactions of ytterbium atoms}},}\ }\href {\doibase
  10.1080/00268970802126608} {\bibfield  {journal} {\bibinfo  {journal}
  {Molecular Physics}\ }\textbf {\bibinfo {volume} {106}},\ \bibinfo {pages}
  {1525--1529} (\bibinfo {year} {2008})}\BibitemShut {NoStop}%
\bibitem [{\citenamefont {Lutz}\ and\ \citenamefont {Hutson}(2016)}]{Lutz2016}%
  \BibitemOpen
  \bibfield  {author} {\bibinfo {author} {\bibfnamefont {J.~J.}\ \bibnamefont
  {Lutz}}\ and\ \bibinfo {author} {\bibfnamefont {J.~M.}\ \bibnamefont
  {Hutson}},\ }\bibfield  {title} {\enquote {\bibinfo {title} {{Deviations from
  Born-Oppenheimer mass scaling in spectroscopy and ultracold molecular
  physics}},}\ }\href {\doibase 10.1016/j.jms.2016.08.007} {\bibfield
  {journal} {\bibinfo  {journal} {Journal of Molecular Spectroscopy}\ }\textbf
  {\bibinfo {volume} {330}},\ \bibinfo {pages} {43--56} (\bibinfo {year}
  {2016})}\BibitemShut {NoStop}%
\bibitem [{\citenamefont {Pachucki}\ and\ \citenamefont
  {Komasa}(2008)}]{Pachucki2008}%
  \BibitemOpen
  \bibfield  {author} {\bibinfo {author} {\bibfnamefont {K.}~\bibnamefont
  {Pachucki}}\ and\ \bibinfo {author} {\bibfnamefont {J.}~\bibnamefont
  {Komasa}},\ }\bibfield  {title} {\enquote {\bibinfo {title} {{Nonadiabatic
  corrections to the wave function and energy}},}\ }\href {\doibase
  10.1063/1.2952517} {\bibfield  {journal} {\bibinfo  {journal} {The Journal of
  Chemical Physics}\ }\textbf {\bibinfo {volume} {129}},\ \bibinfo {pages}
  {034102} (\bibinfo {year} {2008})}\BibitemShut {NoStop}%
\bibitem [{\citenamefont {Feldman}\ and\ \citenamefont
  {Cousins}(1998)}]{Feldman1998}%
  \BibitemOpen
  \bibfield  {author} {\bibinfo {author} {\bibfnamefont {G.~J.}\ \bibnamefont
  {Feldman}}\ and\ \bibinfo {author} {\bibfnamefont {R.~D.}\ \bibnamefont
  {Cousins}},\ }\bibfield  {title} {\enquote {\bibinfo {title} {{Unified
  approach to the classical statistical analysis of small signals}},}\ }\href
  {\doibase 10.1103/PhysRevD.57.3873} {\bibfield  {journal} {\bibinfo
  {journal} {Physical Review D}\ }\textbf {\bibinfo {volume} {57}},\ \bibinfo
  {pages} {3873--3889} (\bibinfo {year} {1998})}\BibitemShut {NoStop}%
\bibitem [{\citenamefont {Balcerzak}\ \emph {et~al.}(2017)\citenamefont
  {Balcerzak}, \citenamefont {Lesiuk},\ and\ \citenamefont
  {Moszynski}}]{Balcerzak2017}%
  \BibitemOpen
  \bibfield  {author} {\bibinfo {author} {\bibfnamefont {J.~G.}\ \bibnamefont
  {Balcerzak}}, \bibinfo {author} {\bibfnamefont {M.}~\bibnamefont {Lesiuk}}, \
  and\ \bibinfo {author} {\bibfnamefont {R.}~\bibnamefont {Moszynski}},\
  }\bibfield  {title} {\enquote {\bibinfo {title} {{Calculation of Araki-Sucher
  correction for many-electron systems}},}\ }\href {\doibase
  10.1103/PhysRevA.96.052510} {\bibfield  {journal} {\bibinfo  {journal}
  {Physical Review A}\ }\textbf {\bibinfo {volume} {96}},\ \bibinfo {pages}
  {052510} (\bibinfo {year} {2017})}\BibitemShut {NoStop}%
\end{thebibliography}%

%\vspace{0.2cm}

\noindent\textbf{Acknowledgments.} We thank T. Zelevinsky, R. Moszyński, P. Żuchowski, K. Enomoto, M. Ando, J. M. Hutson, and E.~Tiemann for useful discussions. We also thank K. Takahashi for his experimental assistance. This work has been partially supported by the Grant-in-Aid for Scientific Research of JSPS (Grants No. 18H05228, No. 18H05405, No. 7H06138), Impulsing Paradigm Changing Through Disruptive Technologies (ImPACT) program, JST CREST (Grant No. JPMJCR1673), and Matsuo Foundation. We acknowledge partial support by Russian Science Foundation Grant No. 17-13-01466.  This research was also partially supported by the COST Action CM1405 MOLIM. We acknowledge support from the National Science Centre (Grant Nos. 2014/13/N/ST2/02591 and 2017/25/B/ST4/01486).  Support has been received from project EMPIR 15SIB03 OC18. This project has received funding from the EMPIR programme co-financed by the Participating States and from the European Union’s Horizon 2020 research and innovation programme. This work is part of an ongoing research program at the National Laboratory FAMO in Toruń, Poland. Calculations have been carried out at the Wroclaw Centre for Networking and Supercomputing (http://www.wcss.pl), Grant No. 353.
%\vspace{0.2cm}

\noindent\textbf{Author Contributions} All authors discussed the results, contributed to the data analysis and worked together on the manuscript.
%\vspace{0.2cm}

\noindent\textbf{Author Information} The authors declare no competing financial interests.
\end{document}